\documentclass[aps, prb,twocolumn,superscriptaddress, nofootinbib, showpacs]{revtex4-2}
\usepackage{dcolumn}

\usepackage{graphicx}% Include figure files
\usepackage{dcolumn}% Align table columns on decimal point

\usepackage{textgreek}
\usepackage{textcomp}
\usepackage{adjustbox}
\usepackage{multirow}

\def\ECA{EuCd$_2$As$_2$}

\def\EBCA{Eu$_{1-x}$Ba$_x$Cd$_2$As$_2$}

\def\Tc{$T_c$}

\def\degrees{$^{\circ}$}

\def\Ptmo{$P\overline{3}m1$}

\begin{document}

\preprint{APS/123-QED}

\title{Single pair of Weyl nodes in the spin-canted structure of EuCd$_2$As$_2$}

%Authors

\author{K.M. Taddei}
\thanks{These authors contributed equally}
\affiliation{Neutron Scattering Division, Oak Ridge National Laboratory, Oak Ridge, TN 37831}
\email[corresponding author ]{taddeikm@ornl.gov}
\author{L. Yin}
\thanks{These authors contributed equally}
\affiliation{Materials Science and Technology Division, Oak Ridge National Laboratory, Oak Ridge, TN 37831}
\author{L.D. Sanjeewa}
\affiliation{Materials Science and Technology Division, Oak Ridge National Laboratory, Oak Ridge, TN 37831}
\author{Y. Li}
\affiliation{Materials Science Division, Argonne National Laboratory, Lemont, IL 60439}
\author{J. Xing}
\affiliation{Materials Science and Technology Division, Oak Ridge National Laboratory, Oak Ridge, TN 37831}
\author{C. dela Cruz}
\affiliation{Neutron Scattering Division, Oak Ridge National Laboratory, Oak Ridge, TN 37831}
\author{D. Phelan}
\affiliation{Materials Science Division, Argonne National Laboratory, Lemont, IL 60439}
\author{A.S. Sefat}
\affiliation{Materials Science and Technology Division, Oak Ridge National Laboratory, Oak Ridge, TN 37831}
\author{D. Parker}
\affiliation{Materials Science and Technology Division, Oak Ridge National Laboratory, Oak Ridge, TN 37831}

\date{\today}

\begin{abstract}
Time reversal symmetry breaking Weyl semimetals are unique among Weyl materials in allowing the minimal number of Weyl points thus offering the clearest signatures of the associated physics. Here we present neutron diffraction, density functional theory and transport measurement results which indicate that \ECA , under ambient field, strain and pressure, is such a material with a single pair of Weyl points. Our work reveals a magnetic structure (magnetic space group $C2'/m'$) with Eu moments pointing along the [210] direction in-plane and canted $\sim$ 30\degrees\ out-of-plane. Density functional theory calculations using this structure show that the observed canting drastically alters the relevant electronic bands, relative to the in-plane order, leading to a single set of well defined Weyl points. Furthermore, we find the canting angle can tune the distance of the Weyl points above the Fermi level, with the smallest distance at low canting angles. Finally, transport measurements of the anomalous Hall Effect and longitudinal magnetoresistance exhibit properties indicative of a chiral anomaly, thus supporting the neutron scattering and DFT results suggesting \ECA\ is close to the ideal situation of the Weyl \lq Hydrogen atom \rq .  
 
\end{abstract}

%\pacs{74.25.Dw, 74.62.Dh, 74.70.Xa, 61.05.fm}

\maketitle


\section{\label{sec:intro}Introduction}

The development of the relativistic quantum mechanical wave equation was a watershed moment in the establishment of quantum mechanics \cite{Dirac1928a,Esposito2012}. With its successful realization came an assortment of newly predicted particles with exotic characteristics such as chirality; broken Lorentz-symmetry; and particle as antiparticle; born at times of \lq beautiful\rq\ simplifications of the Dirac equation - usually without direct physical motivation \cite{Vafek2014}. Perhaps unsurprisingly, many of these particles have never been found  in experimental high energy physics, yet several of them have manifested in condensed matter settings as quasi-particles and excitations. 

However, whereas in the Dirac equation many of these solutions were found through simple manipulations, in condensed matter systems they face a problem of careful tuning, requiring a band structure such that the quasiparticles and low energy exicitations may be described sufficiently well by the relevant formula. In general this requires a specific orbital character to the relevant bands, pre-determined crystal symmetries and a careful tuning of the feature to energy scales relevant for the material's energenetics in order for the solutions to be approached and the associated physics to be detected and available for use in existing and new technologies \cite{Herring1937,Goikoetxea2020,Hasan2017}. Consequently, few ideal realizations which allow full exploration of the resultant physics of Weyl, Axion, Majorana or other such particles, have been found despite general observations that strongly suggest their presence. 

For instance, Weyl Fermions (massless chiral spin-1/2 particles) were first predicted as quasi-particles in the 1930s but not convincingly observed as intrinsic to a material until 2015 and then only via fundamental signatures as quasiparticles (such as Fermi arcs) rather than in their more practically useful bulk transport \cite{Herring1937,Lv2015a,Lv2015b,Liu2019,Chang2018,Soh2019y, Shekhar2018,Xu2020,Loganayagam2012}. This is at least partially due to the requirements needed to create Weyl Fermion-like quasi-particles which necessitate a semi-metal with a fourfold degenerate linearly dispersing band crossing (i.e. a Dirac point) at the Fermi level ($E_F$) whose spin degeneracy is split by breaking either time reversal (TRS) or inversion symmetry (IS) and for which there is preferably no other band at $E_F$, trivial or otherwise.

\begin{figure*}
	\includegraphics[width=\textwidth]{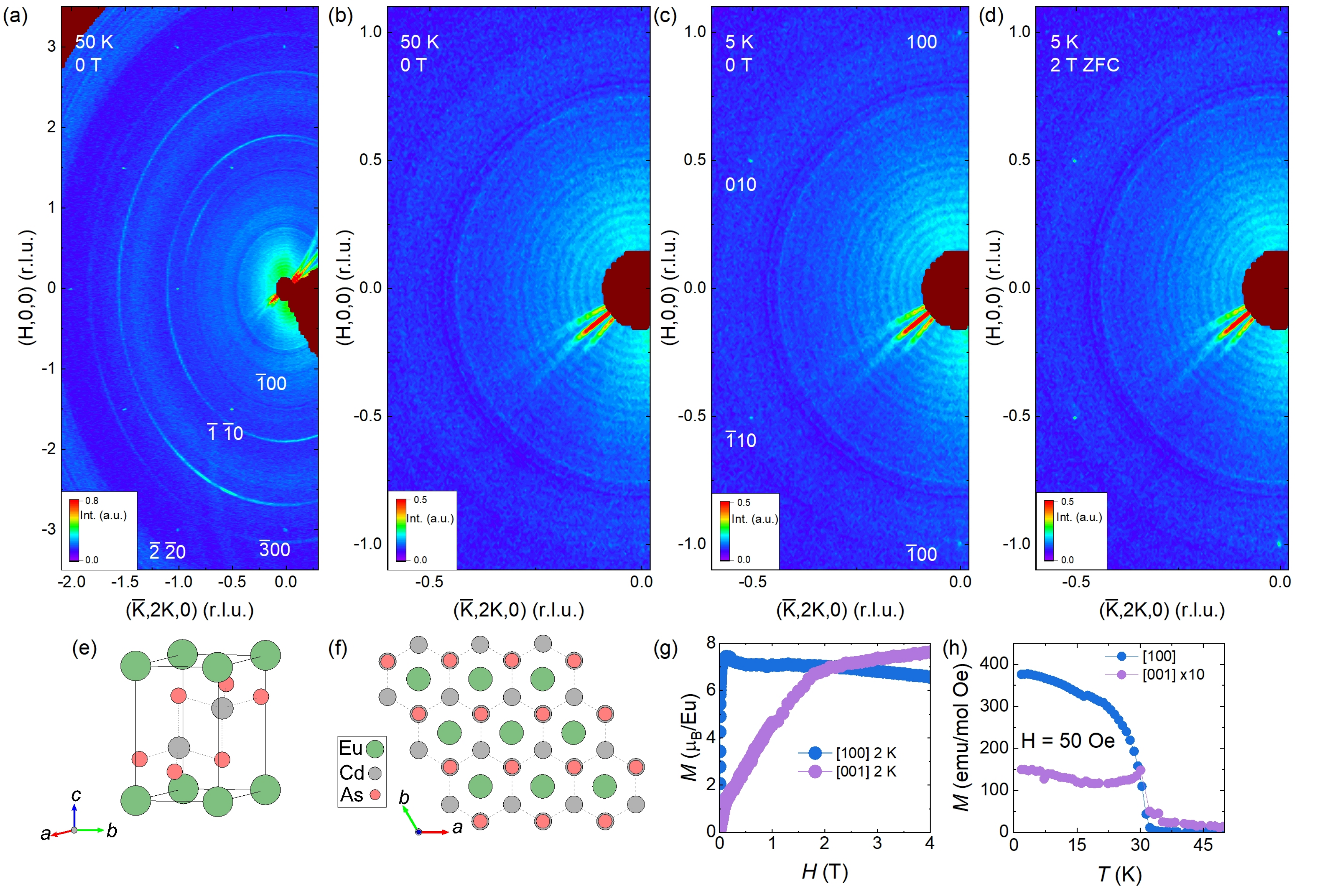}
	\caption{\label{fig:one}  Neutron diffraction patterns of the $HK0$ plane for an isotopic \ECA\ single crystal over a broad range of $H$ and $K$ for data collected at (a) 50 K, 0 T. Zoomed in regions of the $HK0$ plane focusing on the \{100\} series of reflections for data collected at (b) 50 K,0 T; (c) 5 K, 0 T; and (d) 5 K, 2 T. (e) \ECA\ crystal structure viewed along an arbitrary direction to show the layered structure. (f) \ECA\ structure viewed along \textit{c} showing the triangular lattice. (g) $M(H)$ curves for $H$ along different crystallographic directions. (h) $M(T)$ curves for $H$ along different crystallographic directions.}	
\end{figure*}

These requirements are highly restrictive, and thus despite the identification of numerous Weyl semi-metals (WSMs), no such ideal material has been reported absent a perturbing field. \cite{Yan2017, Hasan2017,Armitage2018} This situation is in part due to the inequivalency between the number of and energy difference between Weyl points (WP) generated by TRS breaking (TRSB) and IS breaking (ISB) and nature's seeming preference for the latter \cite{Hasan2017}. As net chirality must always be zero, WPs always appear in pairs with opposite chirality. For an ISB system where TRS is present, a set of WPs at momentum $\textbf{k}$ and $-\textbf{k}$ will have the same chirality and so another pair with the opposite chirality must exist leaving a minimum number of four WPs not constrained to be at the same energy. On the other hand, when TRS is broken while maintaining IS WPs at $\textbf{k}$ and $-\textbf{k}$ must have opposite chiralities and are also forced by symmetry to have the same energy, leaving a simplified system dubbed the \lq Hydrogen atom\rq\ of WSM with the minimum possible of WPs \cite{Hasan2017,Vafek2014}.

Though originally  studied as a potential TRSB Dirac semi-metal, \ECA\ has recently been suggested as a potential ideal WSM - if found to order with the right magnetic space group \cite{Hua2018, Soh2019}. Vexingly, the considerable neutron absorption cross-sections of Cd and Eu have left the magnetic structure effectively unsolved leaving the exact nature of the band structure unknown, an ambiguity which has been further complicated by recent reports of a synthesis dependence to the magnetic structure \cite{Rahn2018,Wang2016, Sanjeewa2020,Jo2020}. None-the-less \ECA\ has evoked topological physics in numerous measurements, with a large anomalous Hall Effect and negative magnetoresistance, an anomalous Nernst effect, a quantum anomalous hall to quantum spin hall insulator transition and linear band dispersions near a band crossing in Angle Resolved Photo Emission Spectroscopy \cite{Hua2018, Sanjeewa2020, Rahn2018, Ma2019, Ma2020, Niu2019, Soh2019}. These results strongly motivate the need to solve \ECA 's magnetic structure and consequently uncover the electronic band structure in the TRSB state. Such information is vital to understanding what type of topological physics is driving the novel bulk behaviors and to uncovering whether \ECA\ may be the much sought after Weyl \lq Hydrogen atom\rq .

In this Letter, using isotopic $^{153}$Eu and $^{116}$Cd we report the zero field and 2 T magnetic structures of FM \ECA\ determined by neutron diffraction, analyze the topology of the resulting electronic band structures and perform transport measurements to identify potential WP driven phenomena. Our data reveal a $\textbf{k}=(0,0,0)$ FM order for both the zero field and 2 T structures. In the former, the Eu moments point along the in-plane [210] direction with a $\sim$ 30\degrees\ out-of-plane canting (MSG $C2'/m'$). For the latter, a \textbf{\textit{c}}-polarized state is found where the moments align with the applied field ($\textbf{H}||\textbf{c}$). Using the canted $C2'/m'$ magnetic structure we performed density functional theory (DFT) calculations and found that the band structure is extraordinarily sensitive to the moment canting, with even small canting angles enhancing the Weyl physics. For $30$\degrees\ canting, we found a single well-defined set of WP with a close proximity to the Fermi level. Finally, using field orientation dependent transport measurements, we find strong signatures of phenomena expected for Weyl physics including a large negative magnetoresistance which is maximized when $\textbf{H}\parallel \textbf{E}$ and a large intrinsic anomalous Hall effect. These results suggest that \ECA\ may be an ideal magnetic WSM with a single set of WPs - the Weyl \lq Hydrogen atom\rq\ - even under ambient field conditions and encourages further work optimizing the canting angle.

Single crystals of \ECA\ were grown following the procedure reported in Ref.~\onlinecite{Sanjeewa2020}. To mitigate the neutron absorption of naturally occurring Eu and Cd, isotopic $^{116}$Cd and $^{153}$Eu (of purity $\sim$99\% ) were used in the reactions and a hexagonally shaped crystal of mass $<$ 1 mg, with a diameter of $\sim$ 1 mm and thickness of $< 0.1$ mm was used for measurements.  Neutron diffraction experiments were carried out on the WAND$^2$ diffractometer of Oak Ridge National Laboratory's High Flux Isotope reactor with incident wavelength 1.48 \AA\ \cite{Calder2018,Frontzek2018}. Symmetry analysis was carried out using the Bilbao Crystallographic Server, SARAh and ISODISTORT \cite{Aroyo2006a,Aroyo2006b,Aroyo2011,Wills2000,Stokes2019,Campbell2006}. Quantitative analyses of the diffraction data were performed using the FullProf software suite \cite{Rodriguez-Carvajal1993}. Crystal structure visualization was performed using VESTA \cite{Momma2011}. First-principles calculations were performed using DFT with spin-orbit coupling as implemented in the Vienna Ab initio Simulation Package (VASP) \cite{Kresse1996,Kresse1996b}. Projector augmented wave pseudo-potentials were applied with the Perdew-Burke-Ernzerhof exchange correlation functional and an energy cutoff of 318 eV \cite{Blochl1994,Kresse1999,Wang2019,Jo2020}.  The magnetoresistance and Hall measurements were carried out in a Quantum Design (QD) Physical Properties Measurment System Dynacool with a Horizontal Rotator option using the standard four-terminal technique. The exact size of the crystal was not measured and so the resistivities are reported in arbitrary units.  For more details on the methods see the supplemental materials (SM) \cite{SM}.    

To better understand the impact of the symmetry of the TRSB state on the topological properties, we start with a discussion of \ECA 's crystal structure (Fig.~\ref{fig:one}(e) and (f)). \ECA\ crystallizes in a trigonal crystal system (space group \Ptmo ). The Eu$^{2+}$ ions occupy the origin of the unit cell creating a triangular sublattice which defines the basal plane (Fig.~\ref{fig:one}(e)). Between these Eu layers is a layer of edge sharing CdAs$_4$ tetrahedra which create a triangluar sublattice (Fig.~\ref{fig:one}(e) and (f)). The \Ptmo\ space group has the $D_{3d}$ symmetry with an implicit $C_3$ rotation axis along the \textbf{\textit{c}}-axis. This rotation provides the symmetry protection for the DP which lie along the $\Gamma$-$A$ line of the Brillouin zone in the paramagnetic state and so its preservation or breaking is vital to considerations of the topology in the magnetically ordered state \cite{Hua2018}. 

In diffraction measurements, a trigonal crystal with the hexagonal lattice will show a 6-fold rotation symmetry in the $HK0$ plane (with the trigonal symmetry appearing when $L\neq 0$) as seen in Fig.~\ref{fig:one} (a). Here the quality of the crystal can be seen in the sharp resolution limited diffraction peaks and we can confirm that the neutron absorption has been mitigated by noting the lack of intensity modulation around any constant scattering vector magnitude ($|\textbf{Q}|$) ring (e.g. the \{110\} series of reflections $(110), (\overline{1}20),(\overline{2}10), (\overline{1}\overline{1}0), (1\overline{2}0),(2\overline{1}0)$ which should have equivalent structure factors in the paramagnetic state $|F_{110}|^2=|F_{\overline{1}20}|^2=|F_{\overline{2}10}|^2=...$). We note that this 6-fold rotation symmetry is generated by the hexagonal lattice setting and therefore the $C_3$ rotation which protects the DP. Any intensity modulation around a constant $|\textbf{Q}|$ ring in the $HK0$ plane is evidence of this symmetry being broken. 

\begin{figure}
	\includegraphics[width=\columnwidth]{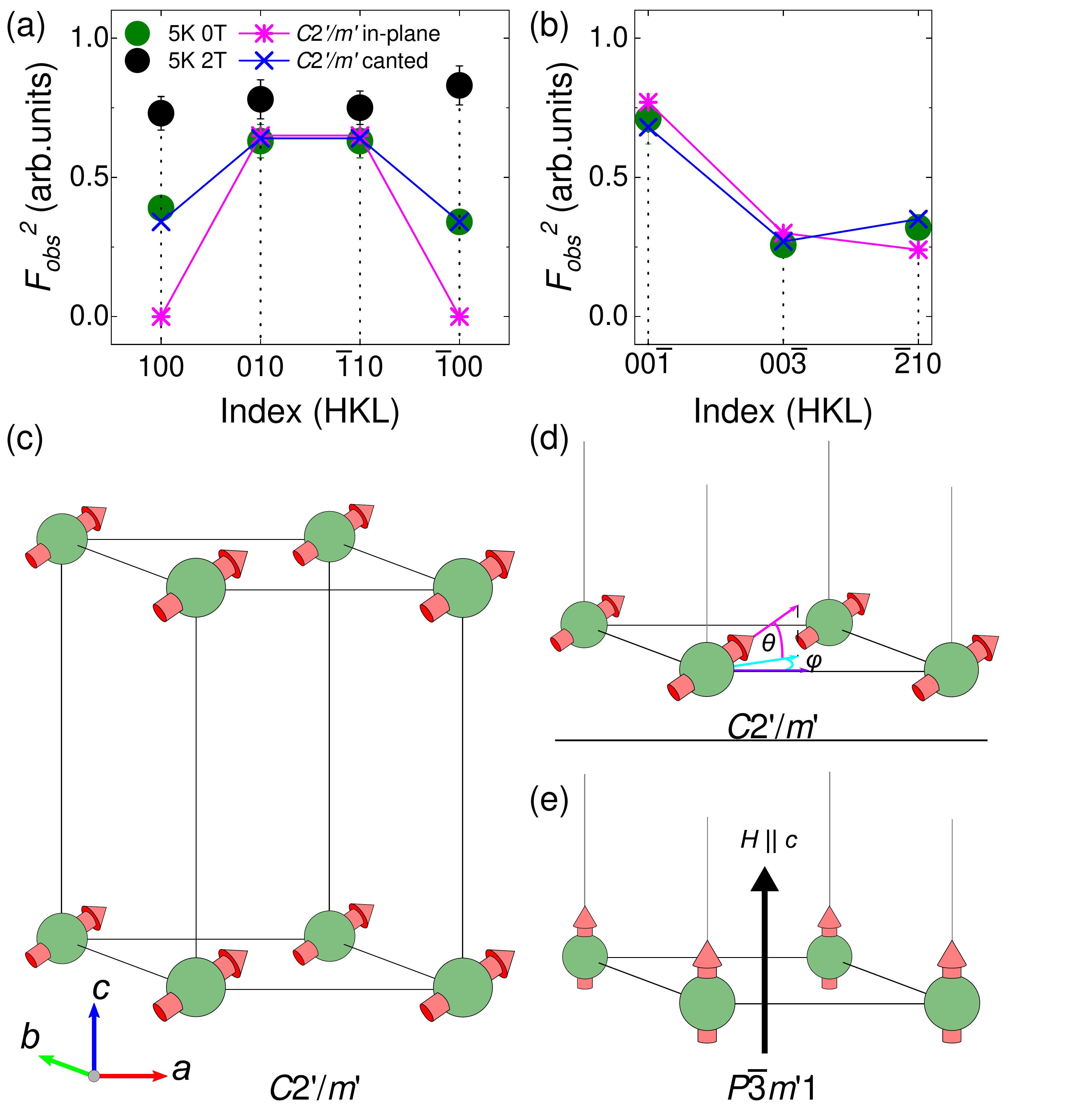}
	\caption{\label{fig:two}  (a) Integrated intensities of the \{100\} series of reflections and (b) of the strongest peaks from the $H2HL$ plane for data collected at 5 K,0 T (green) and 5 K,2 T (black). Plotted with lines are the calculated intensities from best fit models of the $C2'/m'$ magnetic structure with no canting (magenta) and canting (blue). (c) Best fit magnetic structure with moments along the [210] direction with 30\degrees\ canting. (d) diagram showing the definition of the canting angles. (e) \textit{\textbf{c}}-polarized structure for the $\textbf{H}\parallel \textbf{c}$ measurement. }	
\end{figure}

We now turn to the magnetic order which onsets at $T_c \sim 30$ K in our samples as shown in Fig.~\ref{fig:one}(h) and reported previously \cite{Jo2020, Sanjeewa2020}. We note that the samples studied here exhibit FM-like behavior in the magnetization curve rather than the AFM reported in early reports on \ECA\ and point to Refs~\onlinecite{Jo2020} and \onlinecite{Sanjeewa2020} for discussions on the significance of this difference \cite{Schellenberg2011, Wang2016a, Rahn2018}. Shown in Fig.~\ref{fig:one}(b) and (c) are slices of the $HK0$ plane focusing on the small scattering vector region for data collected above (50 K) and below (5 K) \Tc\  respectively. At 50 K no peaks are seen in this region; however, at 5 K intensity is observed at positions indexed by the \{100\} $HKL$ series indicies. While in the \Ptmo\ symmetry these \{100\} reflections are not forbidden, in \ECA\ their structure factors are vanishingly small and so we can identify these peaks as belonging to a magnetic structure which does not break the translational symmetry of the unit cell (i.e. ordering vector of $\textbf{k}=(0,0,0)$). We note that no additional new reflections are seen at fractional coordinates including along the \textit{L} direction. 

Using group theory and representational analysis we considered all allowed subgroups of \Ptmo\ for a $\textbf{k}=(0,0,0)$ which produced unique magnetic structures. Of the four thus obtained MSG the only structure which maintains the $C_3$ symmetry is a \textit{\textbf{c}}-polarized state with MSG $P\overline{3}m'1$ (Fig.~\ref{fig:two}(e)).	The remaining three all significantly lower the space group symmetry with two monoclinic models $C2/m$ and $C2'/m'$ and a low symmetry triclinic $P\overline{1}$ model. The former two of these have constraints on the moment direction with $C2/m$ locking the magnetic moment to the \textit{\textbf{b}}-axis ($\textbf{M}=(0,M_y,0)$) and $C2'/m'$ enforcing the in-plane moment to be along the [210] ($\textbf{M}=(2M_y,M_y,M_z)$) direction while allowing the moment to have a \textit{\textbf{c}}-axis component (Fig.~\ref{fig:two}(c) and (d)) . Meanwhile the $P\overline{1}$ MSG puts no constraints on the moment direction ($\textbf{M}=(M_x,M_y,M_z)$) and can consequently recreate any of the other models.

As magnetic neutron scattering is only sensitive to the moment component perpendicular to $\textbf{Q}$, the relative intensities of peaks in a series of constant $|\textbf{Q}|$ reflections allows for a quick discrimination between these models, for example in the $P\overline{3}m'1$ model all peaks in the \{100\} series are equivalent, while for $C2/m$ the $(100)$ and ($\overline{1}00$) peaks will have higher intensity than the remaining peaks. Knowing this, we can immediately eliminate these two models. Considering the remaining two possible structures, weaker $(100)$ and $(\overline{1}00)$ reflections are consistent with a moment along the [210] direction as described by $C2'/m'$ and thus favor that model (Fig.~\ref{fig:two}(c)). 

To quantify the moment size and direction as well as more rigorously test the various models, we performed Rietveld refinements using all four magnetic structures. To better constrain the moment direction, a second set of neutron diffraction data was collected in a perpendicular plane with access to peaks with non-zero $L$ and integrated for use in the refinements (Fig.~\ref{fig:two} (b) and the SM \onlinecite{SM}). In this analysis we found  the $C2'/m'$ structure to definitively produce the best fit of the observed intensities. The obtained structure has a total moment of 6.7(4) $\mu_B/$Eu with a canting angle of $\theta \sim 30$\degrees (Fig.~\ref{fig:two}(c)).  The calculated intensities from the model are shown together with the strongest magnetic peaks of the two data sets in Fig.~\ref{fig:two}(a) and (b) demonstrating the good agreement of the model with the data. 
   
\begin{figure}
	\includegraphics[width=\columnwidth]{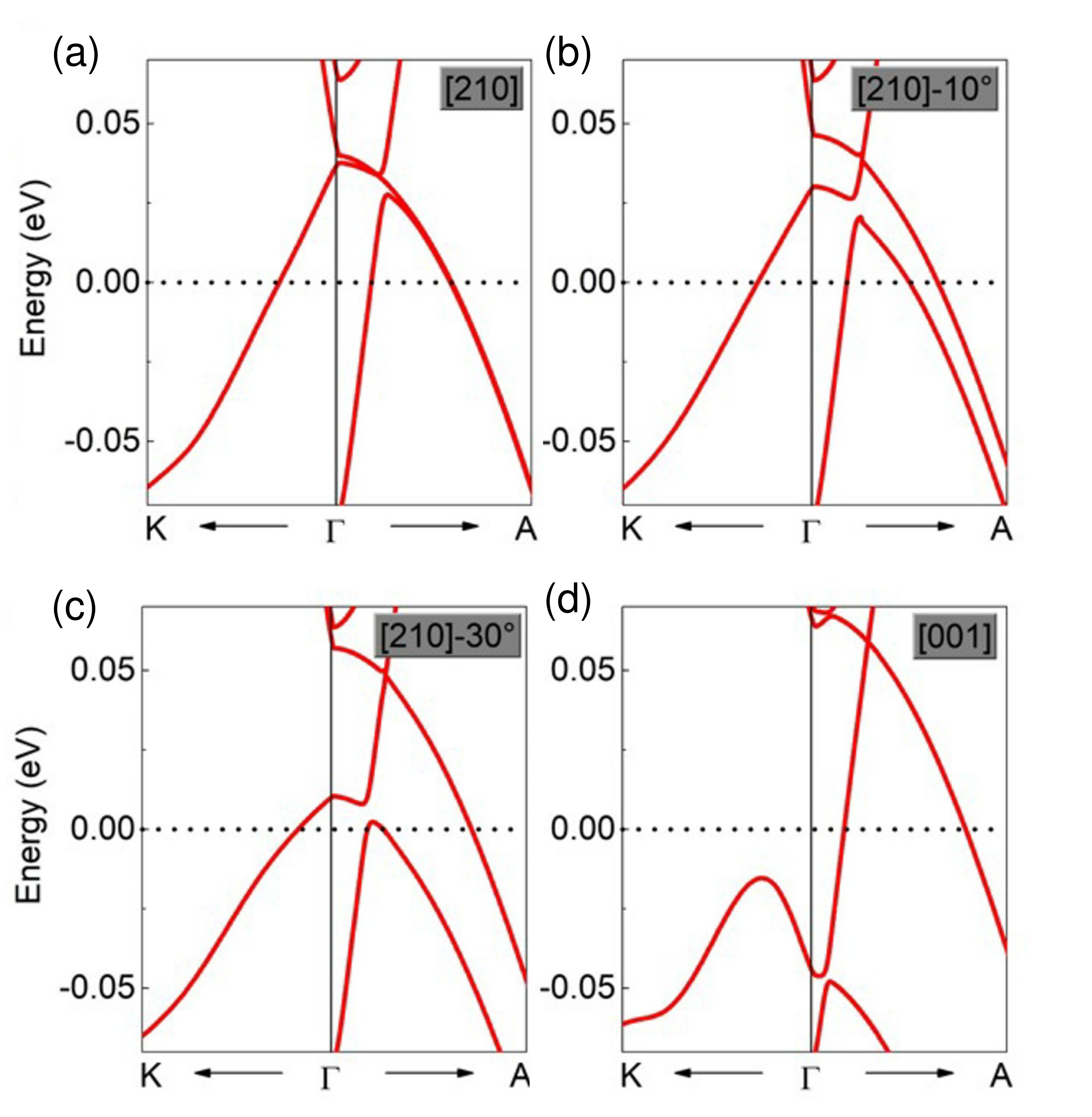}
	\caption{\label{fig:three}  The band structures of ferromagnetic \ECA\ zoomed into the location of the WP for the magnetic moment along the (a) [210] (b) [210]+10\degrees\ canting , (c) [210]+30\degrees\ canting, and (f) [001] directions. The canting angle is defined with respect to the \textit{ab}-plane).}	
\end{figure}

Though allowed in $C2'/m'$, this canting is unexpected and so worth carefully checking. In Fig.~\ref{fig:two}(a) and (b) we show the calculated intensities from best fit models for the structure with a purely in-plane moment and for a canted structure. The purely in-plane structure is unable to reproduce the observed intensity on the (100) and ($\overline{1}00$) Friedel pair which for the [210] moment direction should be zero. However, canting the moment alleviates this and produces an excellent agreement with the observed intensities. We also considered the possibility of magnetic domains with 120\degrees\ rotations of the moment direction, however modeling with domain structures both introduced more refinable parameters and resulted in less satisfactory fits and so the model using a single domain with canting was used for the final refinements (we note that a single domain is consistent with the effective field cooling due to remnant fields in the cryomagnet see SM for details)\cite{SM}. 
  
Briefly, we now consider the data collected under a 2 T applied field (Fig.~\ref{fig:one}(d)). In these data, the intensity on the \{100\} series does not modulate suggesting the moment points purely along the \textit{\textbf{c}}-axis (Fig.~\ref{fig:three}(a). This indicates a metamagnetic transition to the \textit{c}-polarized $P\overline{3}m'1$ MSG as has been previously suggested. We label this as a metamagnetic transition rather than a spin-reorientation within the $C2'/m'$ due to the suggestion of a phase transition in previous studies \cite{Sanjeewa2020,Jo2020}. 

\begin{figure}
	\includegraphics[width=\columnwidth]{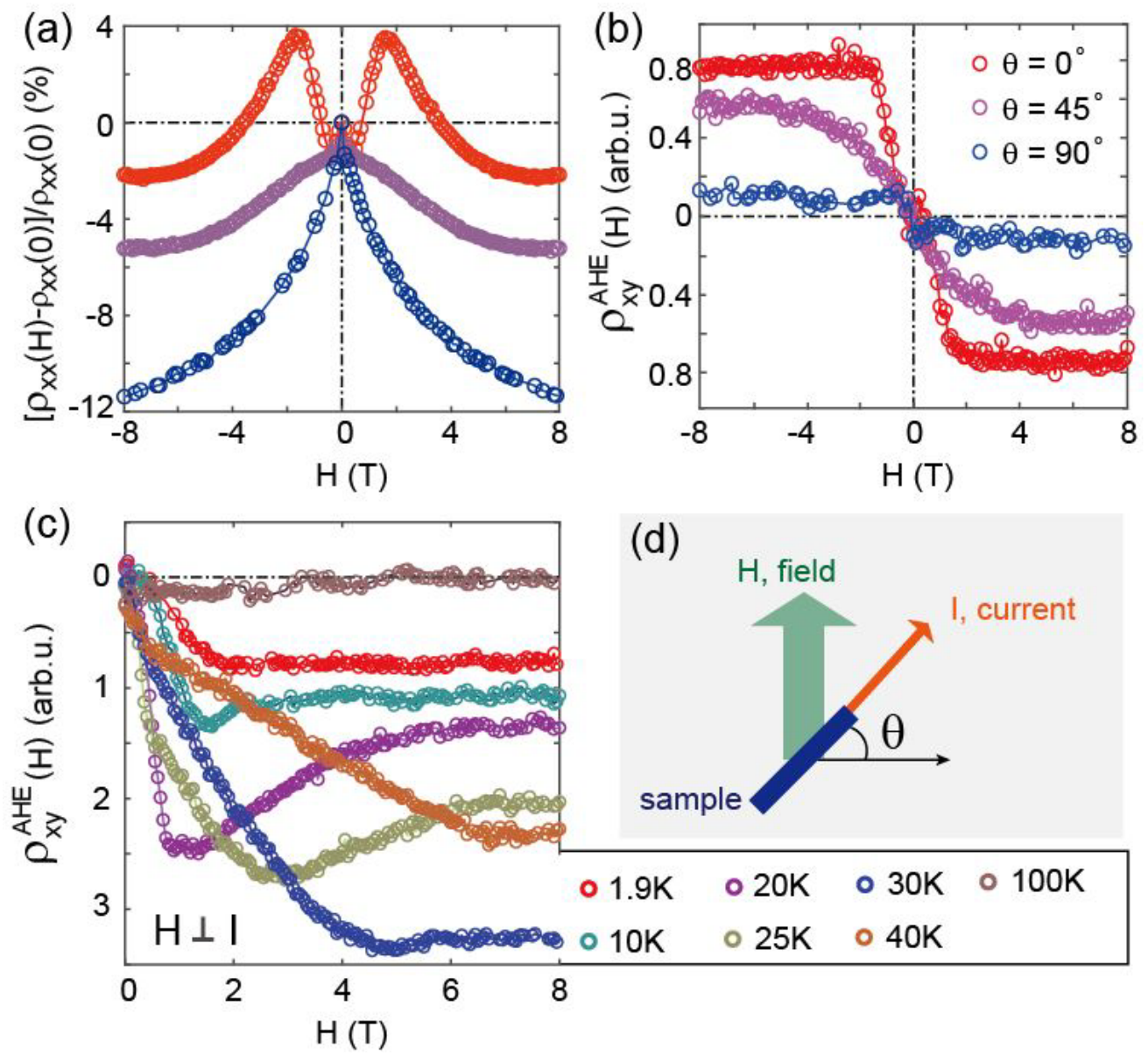}
	\caption{\label{fig:four}  (a) Magnetoresistance for a FM \ECA\ crystal with three different orientations of $\textbf{H}$ and $\textbf{E}$ (b) transverse resistance ($\rho_{xy}$) as the Anomalous Hall Effect for the same three orientations as (a). (c) $\rho_{xy}$ collected in the cononical geometry for several different temperatures. (d) diagram of the MR and AHE measurements showing the geometry of our measurements . }	
\end{figure}

To predict the topological properties associated with these magnetic structures, we performed DFT calculations of the electronic band structure for the purely in-plane [210] direction, the [001] direction as well as for several canting angles in between (Fig.~\ref{fig:three}.) Starting with the [001] structure (Fig.~\ref{fig:four}(a) and (g)), our results are consistent with previous calculations showing a single pair of WPs $\sim$0.06 eV above the Fermi energy ($E_F$) along the $\Gamma$-A direction in the Brillouin zone \cite{Jo2020,Ma2019}.  

For the purely in-plane [210] order (Fig.~\ref{fig:three} (b)), the band structure looks largely similar, except for along the $\Gamma$-A line. Here we still find a WP which is actually closer to $E_F$ but, slightly off of the $\Gamma$-A line (as was reported for the [100] structure) \cite{Jo2020,Ma2019}. This demonstrates the significance of the broken $C_3$ rotation axis in the [210] in-plane order which pulls the WPs off of the $\Gamma -A$ axis. Additionally, the previously steep valence band between the WP and $\Gamma$ becomes nearly flat creating a less ideal situation for the isolation of Weyl physics. 

As the moment cants out of the \textit{ab}-plane, we find that this flat band quickly becomes more dispersive and the band structure more conducive to Weyl physics. For a minor 10\degrees\ canting (Fig.~\ref{fig:three}(d)) the dispersion in the immediate vicinity of the WP already closely resembles the [001] structure. Furthermore, an extensive search of the Brillouin zone revealed that once canted only a single pair of WPs is realized (see SM) \cite{SM}. Looking at the position of the WP relative to $E_F$, we find that the [210] structure is closest at $\sim$ 0.03 eV above the $E_F$. As the canting is increased the WP moves up in energy but always resides lower than the $\sim$0.06 eV of the [001] structure. Together these results indicate that \ECA\ should be a near ideal WSM even under ambient field conditions with a single pair of WPs close to the $E_F$.

The results of canting on the electronic structure are compelling. Though the [001] state has been thought the preferred structure for realizing Weyl physics in \ECA\ our results suggest otherwise. As shown, one does not need to stabilize the fully \textit{\textbf{c}}-polarized state to realize a single set of WPs - this is achieved in the canted [210] structure which creates a situation similar to the [001] band structure while also moving the WP closer to the $E_F$. This indicates that the ideal configuration is close to the realized zero-field ground state magnetic structure without any need for external tuning.

Such topological features in the band structure should give rise to strong signatures in transport properties with a set of WPs causing a chiral anomaly and therefore a large longitudinal negative magnetoresistance (nMR) and a large intrinsic anomalous Hall Effect (AHE) \cite{Son2013,Burkov2014,Burkov2014c,Dos2016,Gorbar2018,Liang2018}. To check for evidence of such effects, as well as further corroborate our model's out-of-plane canting we performed angle dependent transport measurements on a single crystal of \ECA (fig.~\ref{fig:four})). By altering the angle ($\theta$) between the applied $\textbf{H}$ and $\textbf{E}$ fields we could discriminate between different origins to the observed effects. % and indirectly corroborate our spin direction based on recent work ref.~\onlinecite{Cao2021} on the AFM compound which inferred moment canting via the dependence of the AHE on $\theta$.  

To start we consider the most obvious signal of a topological band structure by looking for an intrinsic AHE, in particular one which is non-linear in the magnetization. This has been reported previously for both AFM and FM \ECA ; nonetheless, in Fig.~\ref{fig:four}(c) we show our crystals exhibit a large non-linear AHE for the traditional $\textbf{H}\perp \textbf{E}$ configuration \cite{Jo2020, Rahn2018, Sanjeewa2020}. Upon cooling to just above \Tc\ (at 40 K) the previously flat $\rho_{xy}$ begins to show the signature kink structure of the AHE, which is maximized at \Tc\ and diminishes at lower temperatures. Below \Tc\ we observe non-monotonic behavior in the AHE indicative of topological effects where $\rho_{xy}$ shows a peak feature at $\sim 1.5$ T. The significance of such behavior has been discussed previously, and so we use this observation to demonstrate that our samples exhibit the signatures of topological transport \cite{Cao2021, Rahn2018}. 

Next we discuss the longitudinal MR (Fig.~\ref{fig:four}(a)), for the chiral anomaly, a large nMR is expected when $\textbf{H}$ is parallel to $\textbf{E}$ (i.e. $\theta = 90$) which should show a minimum when rotated such that $\textbf{H}\perp \textbf{E}$ \cite{Hirschberger2016,Liu2018}. This gives a way to discriminate between nMR due to topological transport and that due to purely magnetic effects (which should not show such a $\theta$ dependence)\cite{Nickel1995}. In Fig.~\ref{fig:four}(a) we show the longitudinal MR measured at three different $\theta$ settings at $T=2$ K for $E$ applied in the $ab$ plane (see diagram in Fig.~\ref{fig:four}(d)). In all cases we see nMR however, when $\theta = 90$\degrees we observe a clear maximum in the magnitude of the nMR with a change from $\sim 2$\%\ for $\theta = 0$\degrees\ to $\sim 12$\%\ for $\theta = 90$\degrees . Such an observation is consistent with topological transport where in the $\textbf{H}\parallel \textbf{E}$ configuration the chiral anomaly leads to charge carries being pumped between the Weyl nodes of opposite chirality leading to an additional term contributing to the conductance \cite{Son2013}.    

In summary, we report that \ECA\ orders in the FM $C2'/m'$ MSG where the Eu moments point along the crystallographic [210] direction with a $\sim 30$\degrees\ out-of-plane canting. DFT analysis of this structure shows that the well defined single pair of WPs previously reported for the field stabilized \textbf{\textit{c}}-polarized structure is achieved by the out-of-plane canting of the zero-field structure. Furthermore, we find that the WPs lie closest to the $E_F$ for the in-plane structure and so the minimal canting possible which leads to a well-defined WP is the ideal configuration for optimizing the Weyl physics. Transport measurements demonstrate the effects of the WPs through showing indirect evidence of a chiral anomaly in both a large AHE and nMR the latter of which is maximized when $\textbf{H}\parallel \textbf{E}$ as expected for the chiral anomaly.  Notably, $C2'/m'$ allows canting by symmetry and so continuous tuning of the canting angle should be possible via small perturbations which tune the magnetic interactions. This result increases the intrigue of our recent report on Ba substitution (i.e. \EBCA ) which suggested small Ba substitutions led to moment canting and enhanced Weyl physics as identified by transport measurements \cite{Sanjeewa2020}. Our results suggest \ECA\ is a magnetic WSM with a single pair of WPs under ambient field conditions - a rare if not singular such example - and present a route forward to optimize the Weyl physics.

\textit{Note:} During the review of this paper, another paper was submitted which studied the effect of canting in AFM \ECA\ and found a maximum of the AHE for a canting angle (in our notation) of 60\degrees\ and argue that the configuration of the WP is optimized for this canting angle \cite{Cao2021}. Accepting this hypothesis, we can check with our FM sample and the $\theta$ dependence of $\rho_{xy}$ for where the AHE is optimized (Fig.~\ref{fig:four}(b)). In our study we find the maximum AHE for the $\textbf{H}\parallel \textbf{E}$ configuration which is coarsely consistent then with our findings of a 30\degrees\ moment canting and the optimization of the WP. 

\begin{acknowledgments}

The research is partly supported by the US DOE, BES, Materials Science and Engineering Division. The part of the research conducted at ORNL’s High Flux Isotope Reactor was sponsored by the Scientific User Facilities Division, Office of Basic Energy Sciences (BES), US Department of
Energy (DOE). This research used resources of the Compute and Data Environment for Science (CADES) at ORNL, which is supported by the Office of Science of the U.S. DOE under contract No. DE-AC05-00OR22725. The magneto-transport measurements, which were performed in the Materials Science Division at Argonne National Laboratory, were supported by the U.S. Department of Energy, Office of Science, Basic Energy Sciences, Materials Science and Engineering Division. 

\end{acknowledgments}

Notice of Copyright This manuscript has been authored by UT-Battelle, LLC under Contract No. DE-AC05-00OR22725 with the U.S. Department of Energy. The United States Government retains and the publisher, by accepting the article for publication, acknowledges that the United States Government retains a non-exclusive, paid-up, irrevocable, world-wide license to publish or reproduce the published form of this manuscript, or allow others to do so, for United States Government purposes. The Department of Energy will provide public access to these results of federally sponsored research in accordance with the DOE Public Access Plan (http://energy.gov/downloads/doe-public-access-plan).

%\bibliography{Ref}


\end{document}